\documentclass[twoside,twocolumn,9pt]{article}
\usepackage[utf8x]{inputenc}
\usepackage[T1]{fontenc}
\usepackage{extsizes}
\usepackage[super,sort&compress,comma]{natbib} 
\usepackage[version=3]{mhchem}
\usepackage[left=1.5cm, right=1.5cm, top=1.785cm, bottom=2.0cm]{geometry}
\usepackage{balance}
\usepackage{mathptmx}
\usepackage{sectsty}
\usepackage{graphicx} 
\usepackage{lastpage}
\usepackage[format=plain,justification=justified,singlelinecheck=false,font={stretch=1.125,small,sf},labelfont=bf,labelsep=space]{caption}
\usepackage{float}
\usepackage{fancyhdr}
\usepackage{fnpos}
\usepackage[english]{babel}
\addto{\captionsenglish}{
}
\usepackage{array}
\usepackage{droidsans}
\usepackage{charter}
\usepackage[usenames,dvipsnames]{xcolor}
\usepackage{setspace}
\usepackage[compact]{titlesec}
\usepackage{hyperref}

\definecolor{cream}{RGB}{222,217,201}

\usepackage{units}
\usepackage{verbatim}

\usepackage{xspace}

\usepackage[textsize=small,textwidth=1.2cm]{todonotes}

\newcommand{\ket}[1]{\ensuremath{  | {#1} \rangle}}

\newcommand{\lits}[1]{Refs.~[\!\!\citenum{#1}]\xspace}

\newcommand{\new}{}

\newcommand{\PESbow}{HBB\xspace}
\newcommand{\PESmar}{BBSM\xspace}

\begin{document}
\pagestyle{fancy}
\thispagestyle{plain}
\fancypagestyle{plain}{
\renewcommand{\headrulewidth}{0pt}
}

\makeFNbottom
\makeatletter
\renewcommand\LARGE{\@setfontsize\LARGE{15pt}{17}}
\renewcommand\Large{\@setfontsize\Large{12pt}{14}}
\renewcommand\large{\@setfontsize\large{10pt}{12}}
\renewcommand\footnotesize{\@setfontsize\footnotesize{7pt}{10}}
\makeatother

\renewcommand{\thefootnote}{\fnsymbol{footnote}}
\renewcommand\footnoterule{\vspace*{1pt}\color{cream}\hrule width 3.5in height 0.4pt \color{black}\vspace*{5pt}} 
\setcounter{secnumdepth}{5}

\makeatletter 
\renewcommand\@biblabel[1]{#1}            
\renewcommand\@makefntext[1]{\noindent\makebox[0pt][r]{\@thefnmark\,}#1}
\makeatother 
\renewcommand{\figurename}{\small{Fig.}~}
\sectionfont{\sffamily\Large}
\subsectionfont{\normalsize}
\subsubsectionfont{\bf}
\setstretch{1.125} \setlength{\skip\footins}{0.8cm}
\setlength{\footnotesep}{0.25cm}
\setlength{\jot}{10pt}
\titlespacing*{\section}{0pt}{4pt}{4pt}
\titlespacing*{\subsection}{0pt}{15pt}{1pt}

\fancyfoot{}
\fancyfoot[LO,RE]{\vspace{-7.1pt}\includegraphics[height=9pt]{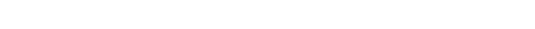}}
\fancyfoot[CO]{\vspace{-7.1pt}\hspace{13.2cm}\includegraphics{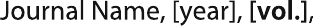}}
\fancyfoot[CE]{\vspace{-7.2pt}\hspace{-14.2cm}\includegraphics{head_foot/RF}}
\fancyfoot[RO]{\footnotesize{\sffamily{1--\pageref{LastPage} ~\textbar  \hspace{2pt}\thepage}}}
\fancyfoot[LE]{\footnotesize{\sffamily{\thepage~\textbar\hspace{3.45cm} 1--\pageref{LastPage}}}}
\fancyhead{}
\renewcommand{\headrulewidth}{0pt} 
\renewcommand{\footrulewidth}{0pt}
\setlength{\arrayrulewidth}{1pt}
\setlength{\columnsep}{6.5mm}
\setlength\bibsep{1pt}

\makeatletter 
\newlength{\figrulesep} 
\setlength{\figrulesep}{0.5\textfloatsep} 

\newcommand{\topfigrule}{\vspace*{-1pt}\noindent{\color{cream}\rule[-\figrulesep]{\columnwidth}{1.5pt}} }

\newcommand{\botfigrule}{\vspace*{-2pt}\noindent{\color{cream}\rule[\figrulesep]{\columnwidth}{1.5pt}} }

\newcommand{\dblfigrule}{\vspace*{-1pt}\noindent{\color{cream}\rule[-\figrulesep]{\textwidth}{1.5pt}} }

\makeatother

\twocolumn[
  \begin{@twocolumnfalse}
{\includegraphics[height=30pt]{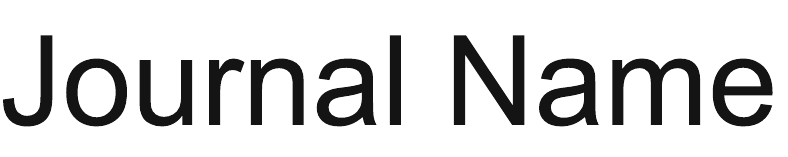}\hfill\raisebox{0pt}[0pt][0pt]{\includegraphics[height=55pt]{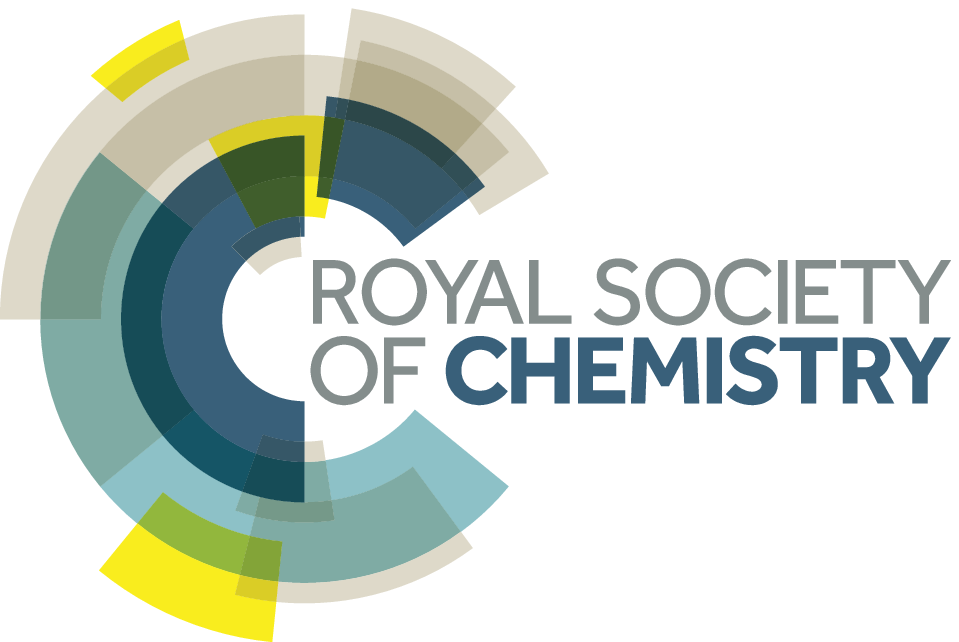}}\\[1ex]
\includegraphics[width=18.5cm]{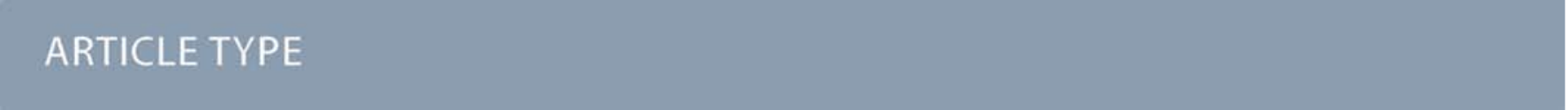}}\par
\vspace{1em}
\sffamily
\begin{tabular}{m{4.5cm} p{13.5cm} }

\includegraphics{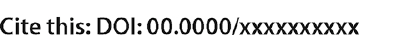} & \noindent\LARGE{\textbf{State-resolved infrared spectrum of the protonated water dimer: Revisiting the characteristic proton transfer doublet peak$^\dag$}} \\\vspace{0.3cm} & \vspace{0.3cm} \\

 & \noindent\large{
Henrik~R.~Larsson,$^{\ast}$\textit{$^{a}$} 
Markus~Schröder,\textit{$^{b}$}
Richard~Beckmann,\textit{$^{c}$}
Fabien~Brieuc,\textit{$^{c,\ddag}$}
    Christoph~Schran,\textit{$^{c,\P}$} Dominik~Marx,\textit{$^{c}$}
and Oriol~Vendrell\textit{$^{b}$}
}\\

\includegraphics{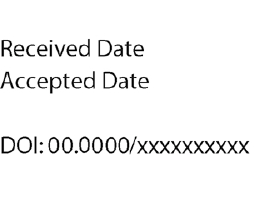} & \noindent\normalsize{The infrared (IR) spectra of protonated water clusters encode precise information on the dynamics and structure of the hydrated proton. 
However, the strong anharmonic coupling 
and quantum effects of these elusive species remain 
puzzling up to the present day. 
Here, we report unequivocal evidence that the interplay between the proton transfer and the water wagging motions in the protonated water dimer (Zundel ion) 
giving rise to the characteristic doublet peak 
is both more complex
and more sensitive to subtle energetic changes than previously thought. 
In particular, hitherto overlooked low-intensity satellite peaks in the experimental spectrum are now unveiled and  mechanistically 
assigned. 
Our findings rely on the comparison of IR~spectra obtained using two highly accurate potential energy surfaces in conjunction with highly accurate
state-resolved quantum simulations.
We demonstrate that these high-accuracy simulations
are important for providing definite assignments of the complex IR signals of fluxional molecules.} \\

\end{tabular}

 \end{@twocolumnfalse} \vspace{0.6cm}

  ]

\renewcommand*\rmdefault{bch}\normalfont\upshape
\rmfamily
\section*{}
\vspace{-1cm}

\footnotetext{\textit{$^{a}$~{azundel\_22a@larsson-research.de}, Department of Chemistry and Biochemistry, University of California, Merced, CA 95343, USA and Division of Chemistry and Chemical Engineering, California Institute of Technology, Pasadena, CA 91125, USA}}
\footnotetext{\textit{$^{b}$~Theoretische Chemie, Physikalisch-Chemisches Institut, Universit\"at Heidelberg, Im Neuenheimer Feld 229, D - 69120 Heidelberg, Germany}}
\footnotetext{\textit{$^{c}$~Lehrstuhl für Theoretische Chemie, Ruhr-Universität Bochum, 44780 Bochum, Germany}}

\footnotetext{\dag~Electronic Supplementary Information (ESI) available: See DOI: 10.1039/D2SC03189B.}

\footnotetext{\ddag~Current Address: Laboratoire Mati\`ere en Conditions Extr\^emes, Universit\'e Paris-Saclay, CEA, DAM, DIF, 91297 Arpajon, France}
\footnotetext{\P~Current Address: Yusuf Hamied Department of Chemistry, University of Cambridge, Lensfield Road, Cambridge, CB2 1EW, UK}

\section{Introduction}
Protonated water clusters play a key role in chemistry and biology, yet their quantum nature continues to puzzle scientists.\cite{GasPhase2003asmis,Infrared2004shin,Spectral2005headrick,Vibrational2014fournier,Spectroscopic2016wolke}
One of the smallest clusters, the protonated water dimer or Zundel ion, \ce{H(H2O)2+}, is particularly important and interesting as it can be  understood as one of the building blocks of acidic water, 
\cite{Nature1999marx,Demystifying2021zeng}
where its 
infrared (IR) 
spectral signatures appear.\cite{Ultrafast2015thamer,Largeamplitude2017dahms,Broadband2018fournier}
Its structure of a proton being ``sandwiched'' by two water units
leads to an unconventional bonding,\cite{Crossover2021dereka,Demystifying2021zeng}
serving as a limiting case of a characteristic pattern 
\cite{Quantum1997tuckerman,Nature1999marx} 
that is observed in a variety of protonated clusters.\cite{Demystifying2021zeng} 
Its IR spectrum is of 
particular importance, as it reveals information about chemical bonding and the quantum nature of molecular vibrations.
However, the IR spectrum  of the Zundel ion
is complicated~-- as such and due to tagging effects.~\cite{Spectral2005headrick,Vibrational2005hammer,Dynamics2007vendrell,Full2007vendrell,Theoretical2010baer}
Unexpectedly, its complexity even increases upon deuteration.\cite{Isotopic2008mccunn,Strong2009vendrell}
Most dominant in the spectrum is the famous 
doublet near $\unit[1000]{cm^{-1}}$, whose vibrational character had not been understood initially.\cite{Vibrational2005hammer}
Quantum dynamics simulations of the Zundel ion allow us to understand the spectrum in full detail.
However, they 
are difficult to perform for the following three reasons: 
(i) its dimensionality with 15 vibrational degrees of freedom, (ii) the non-obvious quantum entanglement of the vibrational zero-order states, and (iii) the fluxional nature of the protons in the Zundel ion with several equivalent minima, connected by only shallow potential barriers.
When combined as in case of the Zundel ion, these difficulties severely challenge
quantum dynamics simulations even today.

To date, the  multiconfiguration time-dependent Hartree (MCTDH) method has been the only one that could fully reproduce the experimental IR spectrum \new{(save for tagging effects)} and it revealed \new{for the first time} the characteristics of the doublet.\cite{Dynamics2007vendrell,Full2007vendrell,Full2009vendrella}
According to MCTDH analysis, 
it 
consists of a Fermi resonance, where low-frequency, low-intensity wagging (pyramidalization) motions of the water units entangle with the proton transfer motion. \new{The analysis has later been confirmed by other simulations.\cite{Theoretical2010baer,Constructing2012dietrick,Reduced2019bertaina}}

While MCTDH wavefunction propagation enables the understanding of the Zundel IR spectrum, the analysis of the vibrational transitions is cumbersome because information on the vibrational structure is only indirectly accessible through observables and low-dimensional probability densities. Further, the spectral resolution of wavefunction propagations is limited by the propagation time and the longer the propagation time, the more complicated the propagated wavefunction becomes. It is desirable to being able to obtain the actual vibrational excited wavefunction that is responsible for a particular peak in the IR spectrum, but this is difficult due to the sheer number of excited eigenstates that are present in such a system of high anharmonicity and dimensionality.

Here, we present new simulations that not only give more accurate vibrational spectra via wavefunction propagations but also reveal the \emph{full eigenstate spectrum} with up to 900 eigenstates to high accuracy for a vibrational energy up to $\unit[\sim\!\!\!1900]{cm^{-1}}$.
Two highly accurate potential energy surfaces (PESs) and dipole moment surfaces (DMSs)
are used, 
namely the pioneering 
permutationally-invariant-polynomial-based
\PESbow surfaces~\cite{Initio2005huang} and 
the more recent \PESmar neural-network-based surfaces,\cite{sch20:88,Infrared2022beckmann}
\new{which so far has not been used for fully, rigorous quantum simulations.}
In the following, by referring to one of the used PESs we imply the usage of the corresponding DMS as well.

Our full-dimensional quantum dynamics
simulations consist of recently introduced vibrational tree tensor network states (TTNS) methodology,\cite{Computing2019larsson} as well as 
new, more accurate and more compact representations of the PES.\cite{Transforming2017schroder,Transforming2020schroder}
The TTNS method is based on the 
the density matrix renormalization group (DMRG),~\cite{dmrg1992white,dmrg1993white}
shares the same wavefunction ansatz as the multilayer (ML) MCTDH method, enables 
both time-independent and time-dependent simulations~\cite{Computing2019larsson,Unifying2016haegeman,Tensor2019schroder} (denoted as ti-TTNS and td-TTNS),
and, most importantly here, the time-independent simulations are 
computationally much less demanding than in ML-MCTDH.\cite{Computing2019larsson}
This allows us to fully identify the nature of each peak in the spectrum by systematically characterising the eigenstates.

The key
finding of this study is that the  interplay between the wagging and the proton motions is more complex than previously thought
in a critical way. 
In particular, we disclose that astonishingly subtle changes in the energetics of the Zundel ion
can lead to very different features in the IR spectrum.
This not only has important implications to PES development and the 
quantum
dynamics of fluxional molecules in general, 
but it will also open the door toward understanding the chemistry of (micro-)solvated species.

Previous MCTDH simulations on the \PESbow PES revealed that the characteristic doublet contains contributions of two states, one with one quantum in the proton transfer motion, and another one with one quantum in the water-water stretch motion and two quanta in the wagging motions. 
Here, in contrast, our simulations on the same PES reveal that there are not two  but three dominant eigenstates that are responsible for the ``doublet'' in the IR spectrum. 
Besides the known  contributions from a state with two quanta in the wagging motions, there is an additional dominant contribution from a state having four quanta
--~only discovered with our novel TTNS-based quantum dynamics method~--
giving overall three dominant peaks in the IR spectrum. In contrast, and in 
accord
with the established knowledge from the MCTDH simulations, the newer \PESmar PES 
displays 
again only 
two peaks for the doublet 
whereas
the state with four quanta in the wagging motion is blueshifted and can 
indeed 
be identified as satellite peak of the doublet in the experimental spectrum,
in support of our assignment.

\section{Setup}

Our benchmark-quality simulations are based on the 
Hamiltonian in polyspherical coordinates set up by Vendrell et al.\cite{Full2007vendrell,Full2009vendrell,Full2009vendrella} 
We fitted both the \PESbow and the \PESmar PESs into a form more suitable for \new{(ML-)}MCTDH and TTNS simulations.
\new{We use the recently developed Monte Carlo  canonical polyadic decomposition (MCCPD) approach for refitting the PESs.\cite{Transforming2017schroder,Transforming2020schroder}
The refitting introduced a mean/root-mean-square error of $\unit[\sim0.2]{cm^{-1}}$/$\unit[\sim11]{cm^{-1}}$.
For comparison, the mean/root-mean-square error of the original cluster expansion of the \PESbow PES used in \lits{Dynamics2007vendrell,Full2007vendrell,Full2009vendrell,Full2009vendrella} is about $\unit[\sim24]{cm^{-1}}$/$\unit[\sim4\cdot10^3]{cm^{-1}}$.\cite{Transforming2020schroder}
To validate the accuracy of both the two PESs and, as additional validation, of the two MCCPDs of the respective PESs, we compare the energies of the PESs from samples of $\sim$2000 configurations with direct, ``gold standard''
 basis-set-extrapolated coupled cluster singles and doubles with perturbative triples (CCSD(T))
electronic structure energies, see Supporting Information for details. With these $\sim$2000 configurations, 
for the \PESbow PES we obtain a mean absolute error  (MAE) of $\unit[203]{cm^{-1}}$, whereas for the \PESmar PES we obtain an improved accuracy with an MAE of $\unit[75]{cm^{-1}}$.
The MCCPDs have essentially the same accuracy.
}

We represent the wavefunction as TTNS, which has the same mathematical structure as the wavefunction used in  the ML-MCTDH method.\cite{Multilayer2003wang,Computing2019larsson}
It differs, however, by the use of more efficient algorithms that are based on a generalization of the density matrix renormalization group (DMRG).\cite{dmrg1992white,dmrg1993white,Computing2019larsson}
For the \PESbow~/ \PESmar PES we computed up to $\sim$900~/ 550 
of the lowest eigenstates, despite the large density of states that is evident in this fluxional cluster.
We verify the eigenstate  ``stick'' spectrum by comparing with the spectrum obtained from wavefunction propagation using the time-dependent DMRG for TTNSs.\cite{Unifying2016haegeman,Tensor2019schroder}
The Fourier-transformed propagation-based spectrum allows us to obtain the full spectrum with broadened peaks,
while the stick spectrum reveals every single detail.
Details of the employed methodology, the numerical parameters, convergence tests, as well as additional results are presented in the Supporting Information.

\section{Results and discussion}

\begin{figure*}[!tbp]
\centering
\includegraphics[width=.8\textwidth]{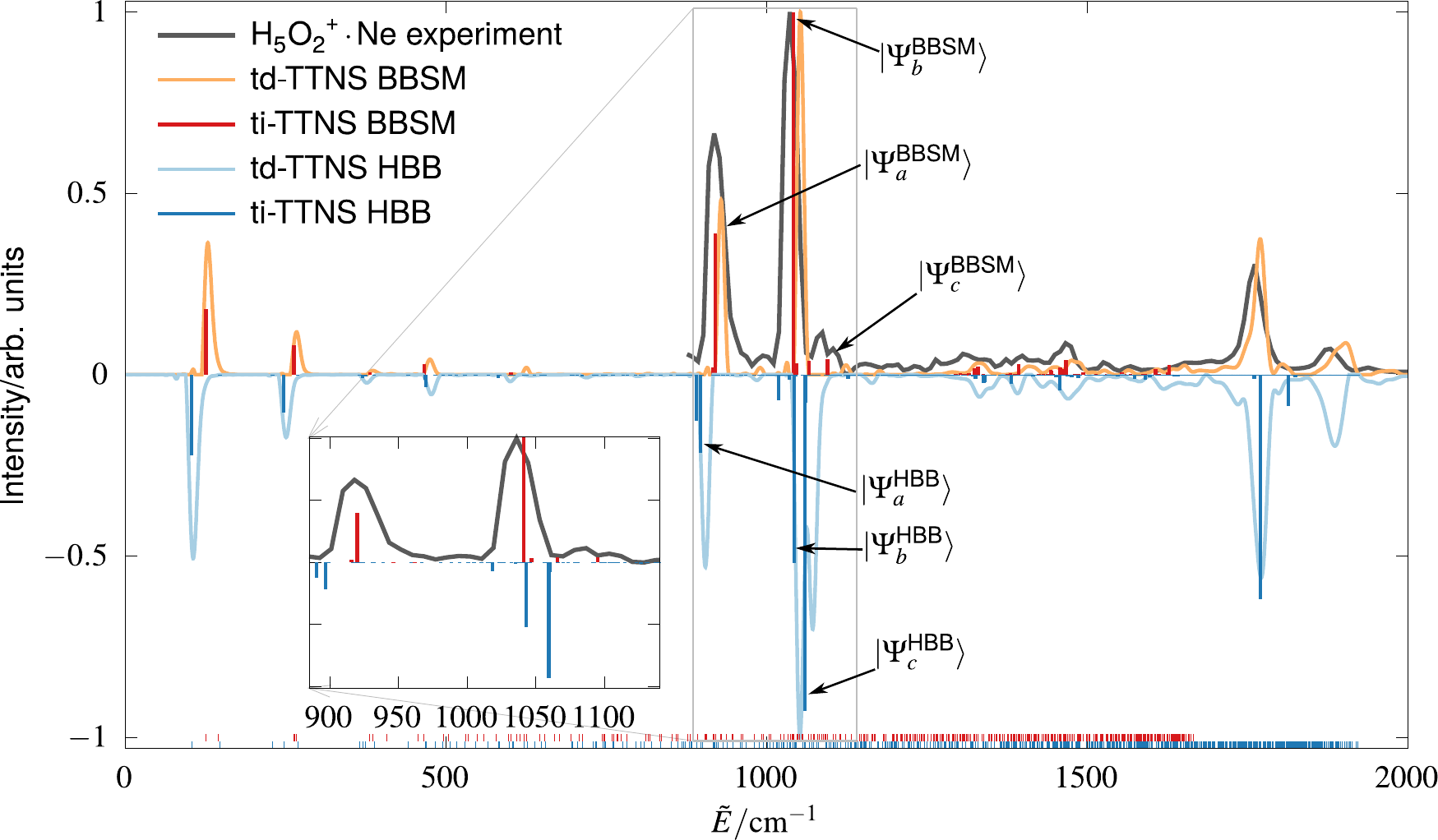}
\caption{
Infrared spectrum of the Zundel ion:
The experimental predissociation spectrum using Neon tagging (gray) \cite{Vibrational2005hammer,Isotopic2008mccunn} and computed spectra using either the 
\PESmar (red) or the \PESbow (blue, negative intensities) potential energy surfaces;
the inset magnified the spectral region in the gray box. 
The computed spectra are either based on time-dependent wavefunction propagations 
(td-TTNS: lines) or based on eigenstate optimizations (ti-TTNS: sticks). 
The wavefunction labels mark the states 
as analyzed in the text.
All eigenstates computed, regardless of intensities, are shown as small lines on the abscissa.
}
  \label{fig:spectrum}
\end{figure*}

\autoref{fig:spectrum} displays the experimental spectrum\cite{Vibrational2005hammer,Isotopic2008mccunn} and the spectra simulated using either of the two used PESs.
For the \PESbow~/ \PESmar PES we show in \autoref{fig:spectrum} up to $\sim$900~/ 550
states up to $\unit[\sim\!\!1900]{cm^{-1}}$ /$\unit[\sim\!\!1700]{cm^{-1}}$;
note
that most of these states are not 
IR-active so that much fewer lines are visible in the spectra. 
The total spectrum, including IR inactive states, 
is shown as small lines on the abscissa. The higher the energy, the higher the  density of states, thus indicating the large 
spectral
complexity of the Zundel ion.
The energies of the IR inactive states differ substantially between the two PESs, which will be discussed in more detail elsewhere.
Compared to the propagation-based spectrum (td-TTNS),  the stick spectrum (ti-TTNS)
has slightly different intensities and the transitions are centered at slightly lower wavenumbers. 
This is because the stick spectrum is not convoluted to account for finite resolution, and because we computed 
the
eigenstate spectrum with higher accuracy than the wavefunction propagation. Convergence tests show that the eigenstates  for levels up to $\unit[\sim\!\!1030]{cm^{-1}}$  are converged to  less than $\unit[1.5]{cm^{-1}}$, much less than the errors of the PES representation (see Supporting Information for further details).
This is the first time that 
so
many states have been computed to such a high accuracy for such a high-dimensional, fluxional cluster.

While both PESs, in particular the newer \PESmar one, yield spectra that overall are in very good agreement with the experimental spectrum, 
a closer look reveals a striking difference between the simulated spectra: 
The most characteristic signal of the IR spectrum, 
the doublet around $\unit[1000]{cm^{-1}}$, 
consists of two dominant peaks on the \PESmar PES
whereas on the \PESbow PES it consists of \emph{three} peaks!
This is in contrast to 
the pioneering
MCTDH simulations on the \PESbow PES.\cite{Dynamics2007vendrell,Full2007vendrell,Strong2009vendrell,Full2009vendrell}
These early MCTDH simulations of such a complicated system were less accurate 
and thus had 
simply
missed the additional peak on the \PESbow PES. 
Indeed, our modern 
ML-MCTDH simulations with improved convergence parameters
also yield three peaks on the \PESbow PES  and overall
are in full agreement with our TTNS simulations.
Comparing the overlaps between the states on each PES 
reveals
that the third peak observed for the \PESbow PES is not missing in the \PESmar PES, but it is 
blueshifted and has a much smaller intensity. 

We focus now on the nature of the three 
dominating vibrational transitions and pinpoint the differences resulting from the 
quality
of the PESs. 
According to \autoref{fig:spectrum},
we label these states as $\Psi_a$, $\Psi_b$, and $\Psi_c$, respectively, see also \autoref{fig:spectrum}.
When needed, we will use superscripts to display which PES has been used for their optimization.
There are two additional, satellite peaks between $\Psi_b^{\text{\PESmar}}$ and $\Psi_c^{\text{\PESmar}}$ that have a similar intensity than $\Psi_c^{\text{\PESmar}}$ and contain a combination of various excitations. 
More details on these and other eigenstates that do not dominate the doublet will be presented elsewhere.

\begin{figure*}[!tbp]
\centering
\includegraphics[width=\textwidth]{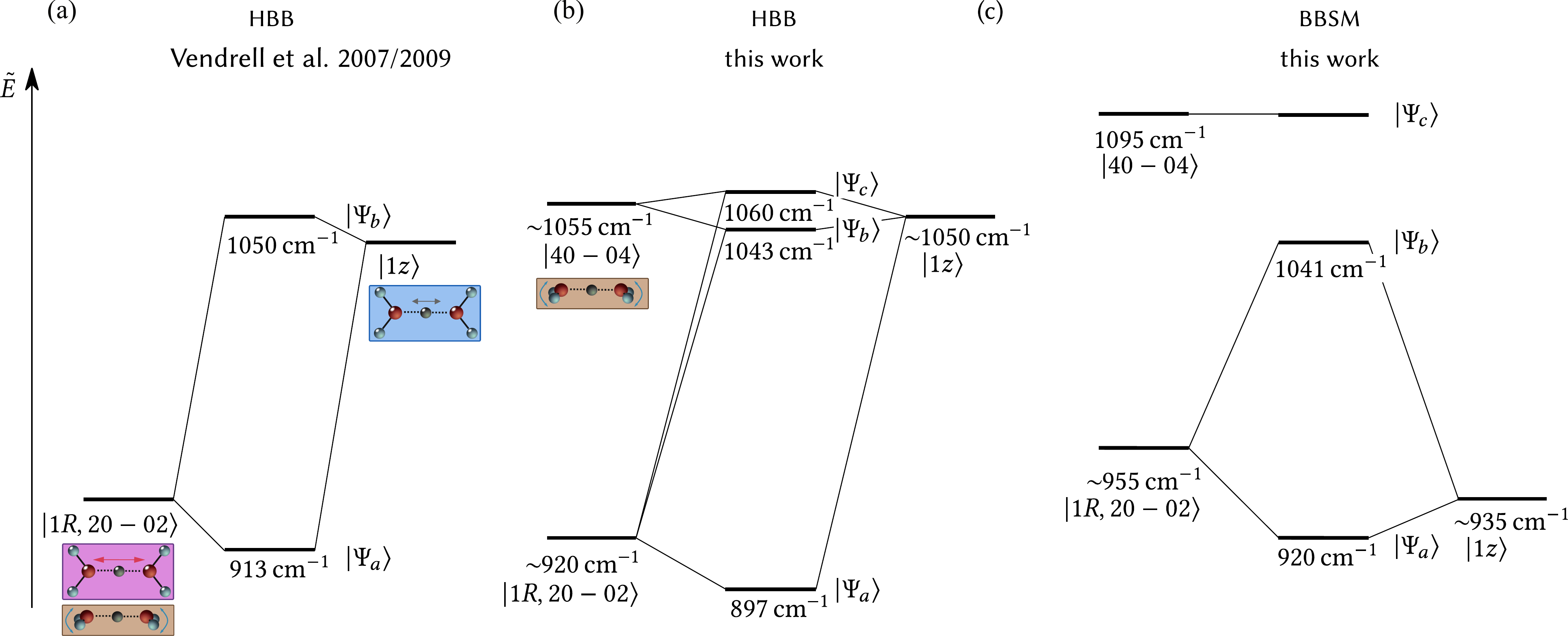}
\caption{Schematics of the coupling scheme of the zero-order states around the doublet in the Zundel infrared spectrum.
    Panel (a) displays the previously identified coupling on the 
pioneering 
\PESbow PES.\cite{Dynamics2007vendrell,Full2009vendrella}
    Panel (b) displays our revised coupling on the same PES.
    Panel (c) displays the coupling on the 
recent
\PESmar PES. 
The insets visualize schematically 
the vibrational excitations of the zero-order states.
    The energies of the zero-order states are  estimated either from the exact $\ket{1R}$ and $\ket{02-20}$/$\ket{04-40}$ states, or from the $\ket{1z, 1\alpha}$ combination state, where $\alpha$ represents the torsion motion.}
  \label{fig:coupling_scheme}
\end{figure*}

Our main findings, namely (i) the nature of the signals dominant in the doublet, (ii) the decomposition of the corresponding wavefunctions into coupling zero-order states, and (iii) their critical dependence on subtle energetic shifts, are 
summarised in \autoref{fig:coupling_scheme}.
Panel (a) displays the coupling -- a Fermi resonance -- identified in the previous MCTDH studies on the \PESbow PES.\cite{Dynamics2007vendrell,Full2009vendrella}
There, the doublet consists of two states that can be described as linear combination (entanglement) of two zero-order states: One zero-order state with one quantum in the  water-water stretch motion ($R$) and two quanta in the two wagging modes ($\ket{02}-\ket{20}$), and another zero-order state with 
one quantum in the proton transfer motion ($z$). 
These two zero-order states are labeled $\ket{1R, 02-20}$ and $\ket{1z}$ respectively.
Panel (b) displays our revised coupling scheme on the \PESbow PES.
There, an \emph{additional} zero-order state $\ket{04-40}$, consisting of \emph{four} quanta in the wagging motion, has a low enough energy that it couples with the two other zero-order states, leading to a more complex quantum resonance and a significant intensity sharing of the $\ket{\Psi_b^\text{\PESbow}}$ and $\ket{\Psi_c^\text{\PESbow}}$.
Panel (c) displays the coupling scheme on the \PESmar PES, which, ironically, is much 
more
similar to that of panel (a)
than to~(b). 
Here, the $\ket{04-40}$ state is 
blueshifted and, therefore, 
does not contribute anymore to the doublet. There is no intensity-sharing and 
$\ket{\Psi_c^\text{\PESmar}}$ 
only appears in the spectrum as one of two satellite peaks
on the blue wing of $\ket{\Psi_b^\text{\PESmar}}$
--~in good agreement with the experimental IR~spectrum as depicted in
the upper panel of \autoref{fig:spectrum}. (The other satellite peak is more complicated and will be discussed elsewhere.)

\begin{figure*}[!tbp]
\centering
  \includegraphics[width=.8\textwidth]{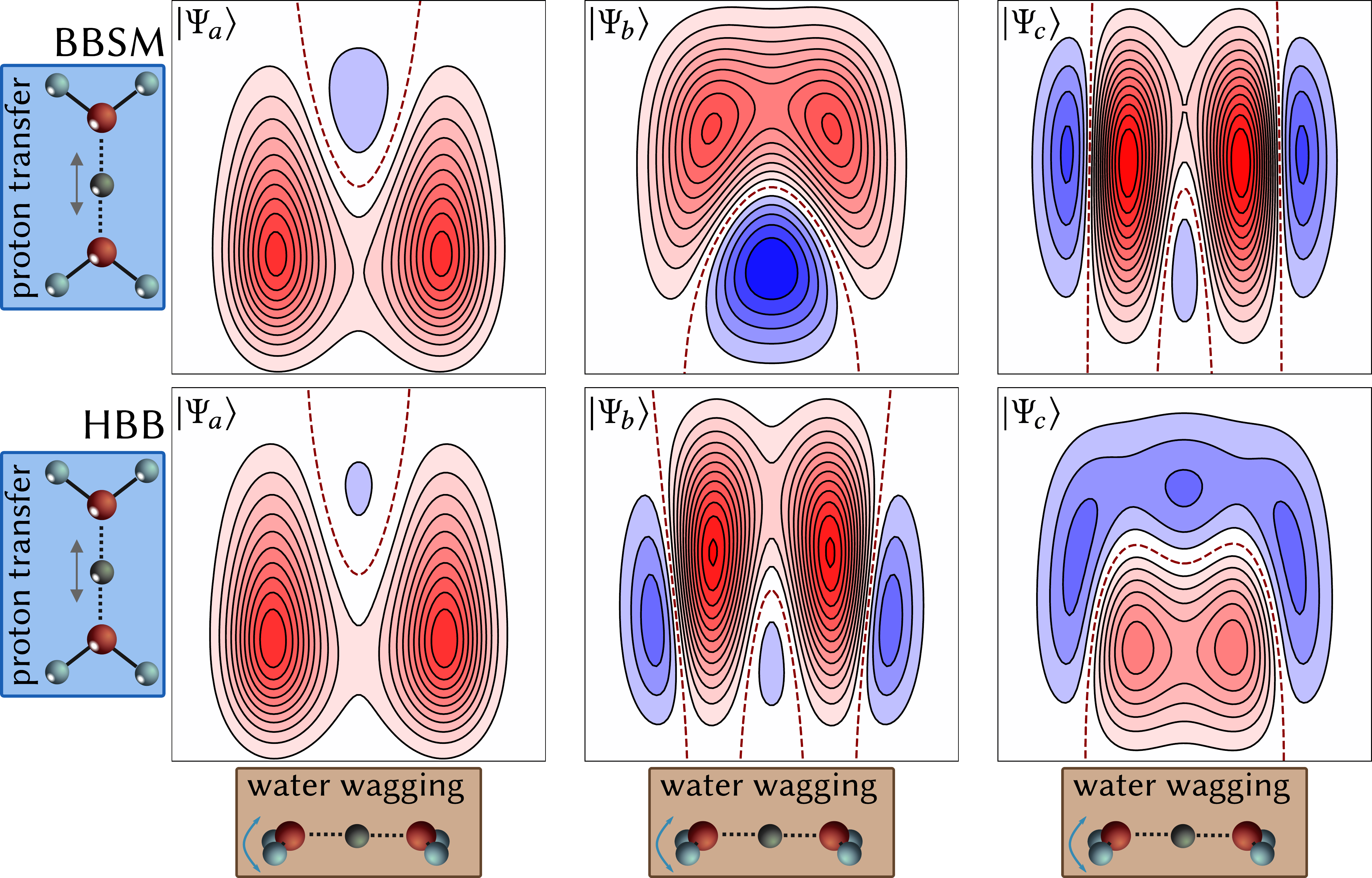}
    \caption{
    Representative cuts of the 
three relevant wavefunctions $\Psi_a$, $\Psi_b$ and $\Psi_c$
corresponding either to 
    the doublet and a satellite peak (\PESmar, upper panels), 
    or  to the triplet (\PESbow PES, lower panels) in the IR spectrum. 
    The abscissa shows the  wagging (pyramidalization) motion of one of the water molecules
whereas the 
ordinate shows the proton transfer motion.
    The red lines denote the zero contours.
        }
  \label{fig:wfs}
\end{figure*}

Let us now 
explain our findings in more detail. Our analysis is based on the time-independent eigenstate simulations, which enable us to directly 
analyze
the wavefunctions responsible for each peak,
thus providing rigorous assignment of IR~peaks to the structural dynamics at 
the level of vibrational motion. 
Moreover, we
confirm this analysis by computing overlaps of the wavefunctions with constructed zero-order states (see Supporting Information).
\autoref{fig:wfs} shows cuts of the three 
most relevant
eigenstates $\Psi_a$, $\Psi_b$, and $\Psi_c$ on the two PESs along one of the wagging motions and the proton transfer motion. 
Each node (zero-crossing) corresponds to one quantum in a particular coordinate. 
The plots of $\Psi_a$ clearly 
indicate
two quanta (nodes) for the wagging motion ($\ket{02-20}$) 
and one quantum for the proton transfer motion ($\ket{1z}$). 
Further inspections reveal an additional quantum for the water-water stretch motion ($\ket{1R}$).
This confirms the assignment shown in \autoref{fig:coupling_scheme}: $\ket{\Psi_a}$ consists of two entangled zero-order states, 
namely
$\ket{02-20; 1R}$ and $\ket{1z}$. 
There is no significant difference between $\Psi_a^{\text{\PESmar}}$ and $\Psi_a^{\text{\PESbow}}$.

\begin{figure}[!tbp]
\centering
\includegraphics[width=\columnwidth]{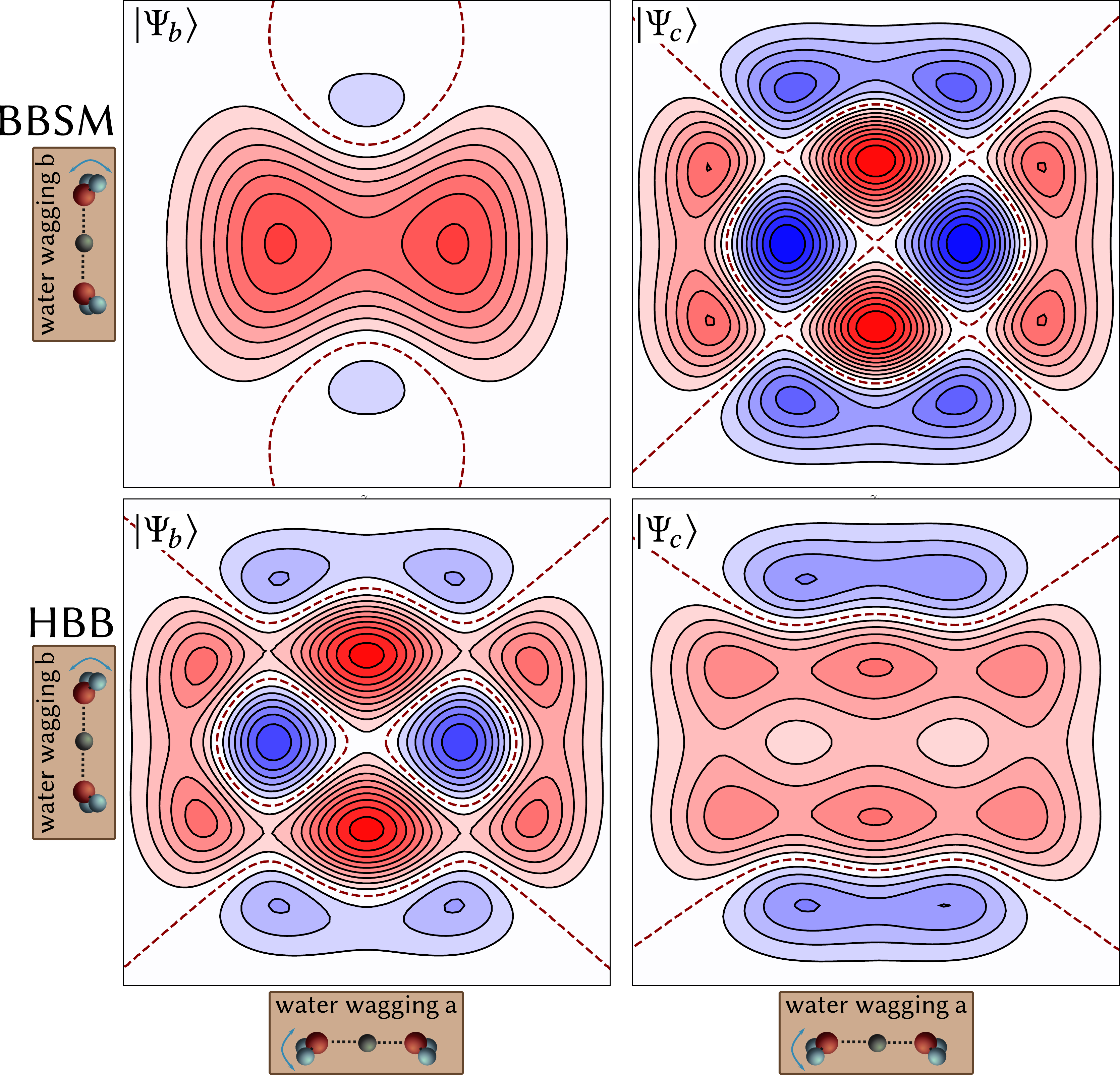}
  \caption{Representative cuts of the two wavefunctions 
$\Psi_b$ and $\Psi_c$ 
along the two wagging motions 
computed with the \PESmar (upper panels) and with the \PESbow (lower panels) PESs.
    The red lines denote the zero contours.
}
  \label{fig:wf_gamma}
\end{figure}

In 
stark
contrast to $\Psi_a$, for $\Psi_b$ and $\Psi_c$ we find significant differences between the \PESmar PES and the \PESbow PES. 
At first sight, $\Psi_b^{\text{\PESmar}}$ looks much more similar to  $\Psi_c^{\text{\PESbow}}$ than to  $\Psi_b^{\text{\PESbow}}$.
We first 
analyze
the states on the \PESmar PES and then compare them with the states on the  \PESbow PES.
$\Psi_b^{\text{\PESmar}}$ has a similar nodal pattern 
as
$\Psi_a$,
as it also displays two quanta along the wagging motion and one quantum along the proton transfer motion.
These 
similarities
are the characteristics of a Fermi resonance. 
$\Psi_c^{\text{\PESmar}}$, however, shows a dominant contribution of not two, but four quanta in the wagging motion.
The rich vibrational structure along the two wagging coordinates is displayed in \autoref{fig:wf_gamma}. This enables us to clearly assign $\Psi_c^{\text{\PESmar}}$ as  $\ket{04-40}$ wagging state.

We now turn to the states on the \PESbow PES.
For $\Psi_b^{\text{\PESbow}}$ and unlike $\Psi_b^{\text{\PESmar}}$, 
in addition to
one quantum in the proton transfer motion, there are not two but four quanta along the wagging motion. \autoref{fig:wf_gamma} clearly shows the similarities to $\Psi_c^{\text{\PESmar}}$. 
Here, however, \autoref{fig:wfs} and additional analysis reveals a more dominant excitation along the proton transfer motion, and we assign $\Psi_b^{\text{\PESbow}}$ as  linear combination of the $\ket{04-40}$ wagging and the $\ket{1z}$ proton transfer zero-order states.
Finally, similar to $\Psi_a$, we identify in $\Psi_c^{\text{\PESbow}}$  
not only dominating contributions  of the
$\ket{02-20; 1R}$ and the $\ket{1z}$ states, but also contributions of the $\ket{04-40}$ wagging state.
Thus, on the \PESbow PES next to the resonance of the $\ket{02-20; 1R}$ and the $\ket{1z}$ state, which alone would lead to two dominant peaks, the  $\ket{1z}$ state also resonates with the $\ket{04-40}$ wagging state.
This coupling creates an additional splitting, leading to overall three peaks in the spectrum, a clear example of intensity sharing.

Why do the three particular zero-order states, $\ket{1R, 02-20}$, $\ket{04-40}$, and $\ket{1z}$, couple on the \PESbow PES and why is the $\ket{04-40}$ not involved in the coupling on the \PESmar PES?
Estimating the energetics of the zero-order states as shown in \autoref{fig:coupling_scheme}
is the clue.
The $\ket{02-20}$ and the $\ket{1R}$ states alone do not have enough energy ($374$ and $\unit[546]{cm^{-1}}$, respectively) to couple with $\ket{1z}$. 
Only a combination of these vibrational excitations leads to a state with similar energy than the $\ket{1z}$ state, which is required for efficient coupling.
Next to the  $\ket{02-20; 1R}$ state, on the \PESbow PES the $\ket{04-40}$ state has an energy that is very close to that of $\ket{1z}$. This leads to 
additional coupling, resulting in $\Psi_b^{\text{\PESbow}}$ and $\Psi_c^{\text{\PESbow}}$. 
While $\ket{04-40}$ and  $\ket{02-20; 1R}$ alone are barely IR active,
large components of the $\ket{1z}$ lead to intensity sharing and thus 
large IR intensities in the spectrum for all three entangled states.

In contrast, on the \PESmar PES the $\ket{04-40}$ state is blueshifted by $\unit[40]{cm^{-1}}$ and the $\ket{1z}$ state is redshifted by $\unit[115]{cm^{-1}}$. This 
disfavors
additional coupling between these states.
Hence, only the $\ket{02-20; 1R}$ couples with $\ket{1z}$, leading to the pair of states, $\Psi_a^{\text{\PESmar}}$ and $\Psi_b^{\text{\PESmar}}$, which form a classical Fermi resonance.
$\Psi_c^{\text{\PESmar}}$ is almost a pure $\ket{04-40}$ state with only minor contributions from $\ket{1z}$ and thus a weak IR intensity.

Which of these two scenarios reflects now the actual situation in experiment? 
While the experimental resolution is not high enough to fully reject the possibility of two states lying under one of the peaks of the doublet, there are two indications that the scenario on the \PESmar PES is more realistic: (1) The peak positions 
obtained from 
the newer \PESmar PES 
are
closer to the experimental spectrum \new{and the PES is much closer to basis-set-extrapolated CCSD(T) energies
as shown in the left panels of Fig.~S3.},
(2) $\Psi_c^{\text{\PESmar}}$ can be attributed to one of two satellite peaks in the experimental spectrum, and (3) another state between $\Psi_b^{\text{\PESmar}}$ and  $\Psi_c^{\text{\PESmar}}$ likely corresponds to the second satellite peak.
We thus 
conclude
that the additional coupling seen in the older \PESbow PES is most likely an artifact that has previously been missed due to less accurate 
computation of the IR spectrum, which is only now disclosed in view of
much improved methodologies.

Based on these findings, it 
is striking 
to demonstrate
that such relatively small changes in the energetics lead to such drastic differences in the entanglement of the eigenstates and in the IR intensities, 
which serve as the experimental observables for these intricate phenomena.
Moreover, we infer that minute 
changes in the environment can lead to even larger energy differences and we anticipate effects similar to those shown here for microsolvated clusters and larger clusters with additional solvation shells.
To give three examples:
(1) deuteration of the Zundel ion significantly alters the energetics of the zero-order states and vastly complicates the IR spectrum.\cite{Strong2009vendrell}
(2) the experimental spectrum of \ce{H5O2+.Ar} \new{(and that of \ce{H5O2+.H2})} is 
strikingly
different to that 
of  \ce{H5O2+.Ne}, 
as it 
displays four and not two peaks around $\unit[1000]{cm^{-1}}$,\cite{Isotopic2008mccunn} indicating a similar complex coupling situation (the \ce{Ar} atom attaches to one of the \ce{OH} units and thus lifts degeneracy\new{; in contrast, the spectra of \ce{H5O2+.Ne} and \ce{H5O2+.He} do not show significant differences and their spectra should be very close to that of the bare \ce{H5O2+}\cite{Communication2014johnson}}).
(3) recent ML-MCTDH simulations on the
solvated hydronium or
Eigen ion (\ce{H9O4+}) reveal the contribution of dozens of eigenstates that dominate the IR activation of the 
\new{hydronium O-H stretch motion and show that the proton vibrations of the Eigen ion can be understood in terms of an embedded Zundel subunit}.\cite{Coupling2022schroder}
\new{
Likewise, recent experimental studies indicate that the dynamics of protonated water clusters can be related to fluctuations of local electrical fields,\cite{Capturing2020yang}
which also appear for solvated \ce{H5O2+}.\cite{Hydrated2016dahmsa,Largeamplitude2017dahms,Decoding2020carpenter}
}

\section{Conclusions}

In conclusion, we report high-accuracy, state-resolved
simulations 
of
the Zundel ion on two PESs with unprecedented accuracy.
Using variational DMRG-like algorithms adapted to tree tensor network states and polyspherical coordinates, we are able to compute up to 900 eigenstates up to $\unit[1900]{cm^{-1}}$ with 
hitherto unreached accuracy for such a complex. 
Obtaining this large number of eigenstates has so far never been accomplished for a 15-dimensional, fluxional system using curvilinear coordinates, and this will serve as a benchmark for future quantum dynamics method developments. 
Only applying this advanced methodology allows us to draw definite conclusions on the quantum dynamics of the Zundel ion, in particular for the most dominant feature of the IR spectrum, namely the doublet around $\unit[1000]{cm^{-1}}$.
Previous assignments are shown to be incomplete due to several limitations 
of the methods available close to two decades ago.
The novel assignment of the doublet around $\unit[1000]{cm^{-1}}$ not
only confirms the Fermi resonance nature of this proton transfer mode, but also 
explains so-far unassigned low-intensity satellite peaks in the experimental spectrum.
In broader terms, our
findings not only highlight the 
striking mechanistic consequences of the 
strong anharmonicity and fluxional character of this challenging but important molecular cluster, but also reveal how subtle effects of the energetics of the zero-order states can lead to qualitative changes 
of the IR spectrum, revealing
fundamental
quantum effects such as resonances, entanglement and intensity sharing.
We are therefore convinced that the same computational approach as unleashed here
for the first time will unveil similar spectral complexity upon assigning 
high-resolution IR spectra
in a multitude of molecular complexes, notably when it comes to 
proton transfer in (micro-)solvation environments.

\new{\section*{Data availability}
Further data is available from the authors upon reasonable request.
}

\new{\section*{Author contributions}
H.R.L.~and O.V.~conceived the study; 
H.R.L.~designed the study, developed the TTNS methodology and performed and analyzed the TTNS simulations;
M.S.~and O.V.~performed the MCCPD fits and additional ML-MCTDH simulations;
R.B., F.B., C.S., and D.M.~created the interface for the BBSM PES, provided samples for the MCCPD procedure, and performed benchmarks for DMC configurations provided by M.S.;
H.R.L.~drafted the manuscript;
all authors contributed to discussing the results and editing the manuscript. 
}

\section*{Conflicts of interest}
There are no conflicts to declare.

\clearpage
\section*{Acknowledgements}
The authors thank 
Professor Joel Bowman for having made
their potential energy surface available.
HRL acknowledges support from the University of California Merced start-up funding during the last part of this work.
HRL acknowledges support from a postdoctoral fellowship from the German Research Foundation (DFG) via grant LA 4442/1-1 during the first part of this work. 
HRL acknowledges computational
time both on the Pinnacles cluster at UC Merced (supported by NSF OAC-2019144) and at the Resnick High Performance Computing Center, a facility supported by the Resnick Sustainability Institute at the California Institute of Technology.
The Bochum work has been
funded by the \textit{Deutsche Forschungsgemeinschaft}
(DFG, German Research Foundation) under Germany's
Excellence Strategy~-- EXC~2033~-- 390677874~-- RESOLV
as well as by the individual DFG~grant \mbox{MA~1547/19} to DM
and 
supported by the ``Center for Solvation Science ZEMOS''
funded by the German Federal Ministry of Education and Research
and by the Ministry of Culture and Research of North Rhine-Westphalia.
RB acknowledges funding from the
\textit{Studienstiftung des deutschen Volkes}
and
CS acknowledges partial financial support from the
\textit{Alexander von Humboldt-Stiftung}.
MS and OV thank the High Performance Computing Center in Stuttgart (HLRS)
under the grant number HDQM\_MCT as well as the bwHPC project of the state
of Baden-W{\"u}rttemberg under grant number bw18K011 for
providing computational resources.

\balance

\providecommand*{\mcitethebibliography}{\thebibliography}
\csname @ifundefined\endcsname{endmcitethebibliography}
{\let\endmcitethebibliography\endthebibliography}{}

\bibliographystyle{rsc}

\begin{mcitethebibliography}{38}
\providecommand*{\natexlab}[1]{#1}
\providecommand*{\mciteSetBstSublistMode}[1]{}
\providecommand*{\mciteSetBstMaxWidthForm}[2]{}
\providecommand*{\mciteBstWouldAddEndPuncttrue}
  {\def\EndOfBibitem{\unskip.}}
\providecommand*{\mciteBstWouldAddEndPunctfalse}
  {\let\EndOfBibitem\relax}
\providecommand*{\mciteSetBstMidEndSepPunct}[3]{}
\providecommand*{\mciteSetBstSublistLabelBeginEnd}[3]{}
\providecommand*{\EndOfBibitem}{}
\mciteSetBstSublistMode{f}
\mciteSetBstMaxWidthForm{subitem}
{(\emph{\alph{mcitesubitemcount}})}
\mciteSetBstSublistLabelBeginEnd{\mcitemaxwidthsubitemform\space}
{\relax}{\relax}

\bibitem[Asmis \emph{et~al.}(2003)Asmis, Pivonka, Santambrogio, Br{\"u}mmer,
  Kaposta, Neumark, and W{\"o}ste]{GasPhase2003asmis}
K.~R. Asmis, N.~L. Pivonka, G.~Santambrogio, M.~Br{\"u}mmer, C.~Kaposta, D.~M.
  Neumark and L.~W{\"o}ste, \emph{Science}, 2003, \textbf{299},
  1375--1377\relax
\mciteBstWouldAddEndPuncttrue
\mciteSetBstMidEndSepPunct{\mcitedefaultmidpunct}
{\mcitedefaultendpunct}{\mcitedefaultseppunct}\relax
\EndOfBibitem
\bibitem[Shin \emph{et~al.}(2004)Shin, Hammer, Diken, Johnson, Walters, Jaeger,
  Duncan, Christie, and Jordan]{Infrared2004shin}
J.-W. Shin, N.~I. Hammer, E.~G. Diken, M.~A. Johnson, R.~S. Walters, T.~D.
  Jaeger, M.~A. Duncan, R.~A. Christie and K.~D. Jordan, \emph{Science}, 2004,
  \textbf{304}, 1137--1140\relax
\mciteBstWouldAddEndPuncttrue
\mciteSetBstMidEndSepPunct{\mcitedefaultmidpunct}
{\mcitedefaultendpunct}{\mcitedefaultseppunct}\relax
\EndOfBibitem
\bibitem[Headrick \emph{et~al.}(2005)Headrick, Diken, Walters, Hammer,
  Christie, Cui, Myshakin, Duncan, Johnson, and Jordan]{Spectral2005headrick}
J.~M. Headrick, E.~G. Diken, R.~S. Walters, N.~I. Hammer, R.~A. Christie,
  J.~Cui, E.~M. Myshakin, M.~A. Duncan, M.~A. Johnson and K.~D. Jordan,
  \emph{Science}, 2005, \textbf{308}, 1765--1769\relax
\mciteBstWouldAddEndPuncttrue
\mciteSetBstMidEndSepPunct{\mcitedefaultmidpunct}
{\mcitedefaultendpunct}{\mcitedefaultseppunct}\relax
\EndOfBibitem
\bibitem[Fournier \emph{et~al.}(2014)Fournier, Johnson, Wolke, Weddle, Wolk,
  and Johnson]{Vibrational2014fournier}
J.~A. Fournier, C.~J. Johnson, C.~T. Wolke, G.~H. Weddle, A.~B. Wolk and M.~A.
  Johnson, \emph{Science}, 2014, \textbf{344}, 1009--1012\relax
\mciteBstWouldAddEndPuncttrue
\mciteSetBstMidEndSepPunct{\mcitedefaultmidpunct}
{\mcitedefaultendpunct}{\mcitedefaultseppunct}\relax
\EndOfBibitem
\bibitem[Wolke \emph{et~al.}(2016)Wolke, Fournier, Dzugan, Fagiani, Odbadrakh,
  Knorke, Jordan, McCoy, Asmis, and Johnson]{Spectroscopic2016wolke}
C.~T. Wolke, J.~A. Fournier, L.~C. Dzugan, M.~R. Fagiani, T.~T. Odbadrakh,
  H.~Knorke, K.~D. Jordan, A.~B. McCoy, K.~R. Asmis and M.~A. Johnson,
  \emph{Science}, 2016, \textbf{354}, 1131--1135\relax
\mciteBstWouldAddEndPuncttrue
\mciteSetBstMidEndSepPunct{\mcitedefaultmidpunct}
{\mcitedefaultendpunct}{\mcitedefaultseppunct}\relax
\EndOfBibitem
\bibitem[Marx \emph{et~al.}(1999)Marx, Tuckerman, Hutter, and
  Parrinello]{Nature1999marx}
D.~Marx, M.~E. Tuckerman, J.~Hutter and M.~Parrinello, \emph{Nature}, 1999,
  \textbf{397}, 601--604\relax
\mciteBstWouldAddEndPuncttrue
\mciteSetBstMidEndSepPunct{\mcitedefaultmidpunct}
{\mcitedefaultendpunct}{\mcitedefaultseppunct}\relax
\EndOfBibitem
\bibitem[Zeng and Johnson(2021)]{Demystifying2021zeng}
H.~J. Zeng and M.~A. Johnson, \emph{Annu. Rev. Phys. Chem.}, 2021, \textbf{72},
  667--691\relax
\mciteBstWouldAddEndPuncttrue
\mciteSetBstMidEndSepPunct{\mcitedefaultmidpunct}
{\mcitedefaultendpunct}{\mcitedefaultseppunct}\relax
\EndOfBibitem
\bibitem[Th{\"a}mer \emph{et~al.}(2015)Th{\"a}mer, De~Marco, Ramasesha, Mandal,
  and Tokmakoff]{Ultrafast2015thamer}
M.~Th{\"a}mer, L.~De~Marco, K.~Ramasesha, A.~Mandal and A.~Tokmakoff,
  \emph{Science}, 2015, \textbf{350}, 78--82\relax
\mciteBstWouldAddEndPuncttrue
\mciteSetBstMidEndSepPunct{\mcitedefaultmidpunct}
{\mcitedefaultendpunct}{\mcitedefaultseppunct}\relax
\EndOfBibitem
\bibitem[Dahms \emph{et~al.}(2017)Dahms, Fingerhut, Nibbering, Pines, and
  Elsaesser]{Largeamplitude2017dahms}
F.~Dahms, B.~P. Fingerhut, E.~T.~J. Nibbering, E.~Pines and T.~Elsaesser,
  \emph{Science}, 2017, \textbf{357}, 491--495\relax
\mciteBstWouldAddEndPuncttrue
\mciteSetBstMidEndSepPunct{\mcitedefaultmidpunct}
{\mcitedefaultendpunct}{\mcitedefaultseppunct}\relax
\EndOfBibitem
\bibitem[Fournier \emph{et~al.}(2018)Fournier, Carpenter, Lewis, and
  Tokmakoff]{Broadband2018fournier}
J.~A. Fournier, W.~B. Carpenter, N.~H.~C. Lewis and A.~Tokmakoff, \emph{Nature
  Chem.}, 2018, \textbf{10}, 932--937\relax
\mciteBstWouldAddEndPuncttrue
\mciteSetBstMidEndSepPunct{\mcitedefaultmidpunct}
{\mcitedefaultendpunct}{\mcitedefaultseppunct}\relax
\EndOfBibitem
\bibitem[Dereka \emph{et~al.}(2021)Dereka, Yu, Lewis, Carpenter, Bowman, and
  Tokmakoff]{Crossover2021dereka}
B.~Dereka, Q.~Yu, N.~H.~C. Lewis, W.~B. Carpenter, J.~M. Bowman and
  A.~Tokmakoff, \emph{Science}, 2021, \textbf{371}, 160--164\relax
\mciteBstWouldAddEndPuncttrue
\mciteSetBstMidEndSepPunct{\mcitedefaultmidpunct}
{\mcitedefaultendpunct}{\mcitedefaultseppunct}\relax
\EndOfBibitem
\bibitem[Tuckerman \emph{et~al.}(1997)Tuckerman, Marx, Klein, and
  Parrinello]{Quantum1997tuckerman}
M.~E. Tuckerman, D.~Marx, M.~L. Klein and M.~Parrinello, \emph{Science}, 1997,
  \textbf{275}, 817--820\relax
\mciteBstWouldAddEndPuncttrue
\mciteSetBstMidEndSepPunct{\mcitedefaultmidpunct}
{\mcitedefaultendpunct}{\mcitedefaultseppunct}\relax
\EndOfBibitem
\bibitem[Hammer \emph{et~al.}(2005)Hammer, Diken, Roscioli, Johnson, Myshakin,
  Jordan, McCoy, Huang, Bowman, and Carter]{Vibrational2005hammer}
N.~I. Hammer, E.~G. Diken, J.~R. Roscioli, M.~A. Johnson, E.~M. Myshakin, K.~D.
  Jordan, A.~B. McCoy, X.~Huang, J.~M. Bowman and S.~Carter, \emph{J. Chem.
  Phys.}, 2005, \textbf{122}, 244301\relax
\mciteBstWouldAddEndPuncttrue
\mciteSetBstMidEndSepPunct{\mcitedefaultmidpunct}
{\mcitedefaultendpunct}{\mcitedefaultseppunct}\relax
\EndOfBibitem
\bibitem[Vendrell \emph{et~al.}(2007)Vendrell, Gatti, and
  Meyer]{Dynamics2007vendrell}
O.~Vendrell, F.~Gatti and H.-D. Meyer, \emph{Angew. Chem. Int. Ed.}, 2007,
  \textbf{46}, 6918--6921\relax
\mciteBstWouldAddEndPuncttrue
\mciteSetBstMidEndSepPunct{\mcitedefaultmidpunct}
{\mcitedefaultendpunct}{\mcitedefaultseppunct}\relax
\EndOfBibitem
\bibitem[Vendrell \emph{et~al.}(2007)Vendrell, Gatti, and
  Meyer]{Full2007vendrell}
O.~Vendrell, F.~Gatti and H.-D. Meyer, \emph{J. Chem. Phys.}, 2007,
  \textbf{127}, 184303\relax
\mciteBstWouldAddEndPuncttrue
\mciteSetBstMidEndSepPunct{\mcitedefaultmidpunct}
{\mcitedefaultendpunct}{\mcitedefaultseppunct}\relax
\EndOfBibitem
\bibitem[Baer \emph{et~al.}(2010)Baer, Marx, and Mathias]{Theoretical2010baer}
M.~Baer, D.~Marx and G.~Mathias, \emph{Angew. Chem. Int. Ed.}, 2010,
  \textbf{49}, 7346--7349\relax
\mciteBstWouldAddEndPuncttrue
\mciteSetBstMidEndSepPunct{\mcitedefaultmidpunct}
{\mcitedefaultendpunct}{\mcitedefaultseppunct}\relax
\EndOfBibitem
\bibitem[McCunn \emph{et~al.}(2008)McCunn, Roscioli, Johnson, and
  McCoy]{Isotopic2008mccunn}
L.~R. McCunn, J.~R. Roscioli, M.~A. Johnson and A.~B. McCoy, \emph{J. Phys.
  Chem. B}, 2008, \textbf{112}, 321--327\relax
\mciteBstWouldAddEndPuncttrue
\mciteSetBstMidEndSepPunct{\mcitedefaultmidpunct}
{\mcitedefaultendpunct}{\mcitedefaultseppunct}\relax
\EndOfBibitem
\bibitem[Vendrell \emph{et~al.}(2009)Vendrell, Gatti, and
  Meyer]{Strong2009vendrell}
O.~Vendrell, F.~Gatti and H.-D. Meyer, \emph{Angew. Chem. Int. Ed.}, 2009,
  \textbf{48}, 352--355\relax
\mciteBstWouldAddEndPuncttrue
\mciteSetBstMidEndSepPunct{\mcitedefaultmidpunct}
{\mcitedefaultendpunct}{\mcitedefaultseppunct}\relax
\EndOfBibitem
\bibitem[Vendrell \emph{et~al.}(2009)Vendrell, Gatti, and
  Meyer]{Full2009vendrella}
O.~Vendrell, F.~Gatti and H.-D. Meyer, \emph{J. Chem. Phys.}, 2009,
  \textbf{131}, 034308\relax
\mciteBstWouldAddEndPuncttrue
\mciteSetBstMidEndSepPunct{\mcitedefaultmidpunct}
{\mcitedefaultendpunct}{\mcitedefaultseppunct}\relax
\EndOfBibitem
\bibitem[Dietrick and Iyengar(2012)]{Constructing2012dietrick}
S.~M. Dietrick and S.~S. Iyengar, \emph{J. Chem. Theory Comput.}, 2012,
  \textbf{8}, 4876--4890\relax
\mciteBstWouldAddEndPuncttrue
\mciteSetBstMidEndSepPunct{\mcitedefaultmidpunct}
{\mcitedefaultendpunct}{\mcitedefaultseppunct}\relax
\EndOfBibitem
\bibitem[Bertaina \emph{et~al.}(2019)Bertaina, Di~Liberto, and
  Ceotto]{Reduced2019bertaina}
G.~Bertaina, G.~Di~Liberto and M.~Ceotto, \emph{J. Chem. Phys.}, 2019,
  \textbf{151}, 114307\relax
\mciteBstWouldAddEndPuncttrue
\mciteSetBstMidEndSepPunct{\mcitedefaultmidpunct}
{\mcitedefaultendpunct}{\mcitedefaultseppunct}\relax
\EndOfBibitem
\bibitem[Huang \emph{et~al.}(2005)Huang, Braams, and Bowman]{Initio2005huang}
X.~Huang, B.~J. Braams and J.~M. Bowman, \emph{J. Chem. Phys.}, 2005,
  \textbf{122}, 044308\relax
\mciteBstWouldAddEndPuncttrue
\mciteSetBstMidEndSepPunct{\mcitedefaultmidpunct}
{\mcitedefaultendpunct}{\mcitedefaultseppunct}\relax
\EndOfBibitem
\bibitem[Schran \emph{et~al.}(2020)Schran, Behler, and Marx]{sch20:88}
C.~Schran, J.~Behler and D.~Marx, \emph{J. Chem. Theory Comput.}, 2020,
  \textbf{16}, 88--99\relax
\mciteBstWouldAddEndPuncttrue
\mciteSetBstMidEndSepPunct{\mcitedefaultmidpunct}
{\mcitedefaultendpunct}{\mcitedefaultseppunct}\relax
\EndOfBibitem
\bibitem[Beckmann \emph{et~al.}(2022)Beckmann, Brieuc, Schran, and
  Marx]{Infrared2022beckmann}
R.~Beckmann, F.~Brieuc, C.~Schran and D.~Marx, \emph{J. Chem. Theory Comput.},
  2022,  in press\relax
\mciteBstWouldAddEndPuncttrue
\mciteSetBstMidEndSepPunct{\mcitedefaultmidpunct}
{\mcitedefaultendpunct}{\mcitedefaultseppunct}\relax
\EndOfBibitem
\bibitem[Larsson(2019)]{Computing2019larsson}
H.~R. Larsson, \emph{J. Chem. Phys.}, 2019, \textbf{151}, 204102\relax
\mciteBstWouldAddEndPuncttrue
\mciteSetBstMidEndSepPunct{\mcitedefaultmidpunct}
{\mcitedefaultendpunct}{\mcitedefaultseppunct}\relax
\EndOfBibitem
\bibitem[Schr{\"o}der and Meyer(2017)]{Transforming2017schroder}
M.~Schr{\"o}der and H.-D. Meyer, \emph{J. Chem. Phys.}, 2017, \textbf{147},
  064105\relax
\mciteBstWouldAddEndPuncttrue
\mciteSetBstMidEndSepPunct{\mcitedefaultmidpunct}
{\mcitedefaultendpunct}{\mcitedefaultseppunct}\relax
\EndOfBibitem
\bibitem[Schr{\"o}der(2020)]{Transforming2020schroder}
M.~Schr{\"o}der, \emph{J. Chem. Phys.}, 2020, \textbf{152}, 024108\relax
\mciteBstWouldAddEndPuncttrue
\mciteSetBstMidEndSepPunct{\mcitedefaultmidpunct}
{\mcitedefaultendpunct}{\mcitedefaultseppunct}\relax
\EndOfBibitem
\bibitem[White(1992)]{dmrg1992white}
S.~R. White, \emph{Phys. Rev. Lett.}, 1992, \textbf{69}, 2863--2866\relax
\mciteBstWouldAddEndPuncttrue
\mciteSetBstMidEndSepPunct{\mcitedefaultmidpunct}
{\mcitedefaultendpunct}{\mcitedefaultseppunct}\relax
\EndOfBibitem
\bibitem[White(1993)]{dmrg1993white}
S.~R. White, \emph{Phys. Rev. B}, 1993, \textbf{48}, 10345--10356\relax
\mciteBstWouldAddEndPuncttrue
\mciteSetBstMidEndSepPunct{\mcitedefaultmidpunct}
{\mcitedefaultendpunct}{\mcitedefaultseppunct}\relax
\EndOfBibitem
\bibitem[Haegeman \emph{et~al.}(2016)Haegeman, Lubich, Oseledets, Vandereycken,
  and Verstraete]{Unifying2016haegeman}
J.~Haegeman, C.~Lubich, I.~Oseledets, B.~Vandereycken and F.~Verstraete,
  \emph{Phys. Rev. B}, 2016, \textbf{94}, 165116\relax
\mciteBstWouldAddEndPuncttrue
\mciteSetBstMidEndSepPunct{\mcitedefaultmidpunct}
{\mcitedefaultendpunct}{\mcitedefaultseppunct}\relax
\EndOfBibitem
\bibitem[Schr{\"o}der \emph{et~al.}(2019)Schr{\"o}der, Turban, Musser, Hine,
  and Chin]{Tensor2019schroder}
F.~A. Y.~N. Schr{\"o}der, D.~H.~P. Turban, A.~J. Musser, N.~D.~M. Hine and
  A.~W. Chin, \emph{Nat. Commun.}, 2019, \textbf{10}, 1062\relax
\mciteBstWouldAddEndPuncttrue
\mciteSetBstMidEndSepPunct{\mcitedefaultmidpunct}
{\mcitedefaultendpunct}{\mcitedefaultseppunct}\relax
\EndOfBibitem
\bibitem[Vendrell \emph{et~al.}(2009)Vendrell, Brill, Gatti, Lauvergnat, and
  Meyer]{Full2009vendrell}
O.~Vendrell, M.~Brill, F.~Gatti, D.~Lauvergnat and H.-D. Meyer, \emph{J. Chem.
  Phys.}, 2009, \textbf{130}, 234305\relax
\mciteBstWouldAddEndPuncttrue
\mciteSetBstMidEndSepPunct{\mcitedefaultmidpunct}
{\mcitedefaultendpunct}{\mcitedefaultseppunct}\relax
\EndOfBibitem
\bibitem[Wang and Thoss(2003)]{Multilayer2003wang}
H.~Wang and M.~Thoss, \emph{J. Chem. Phys.}, 2003, \textbf{119},
  1289--1299\relax
\mciteBstWouldAddEndPuncttrue
\mciteSetBstMidEndSepPunct{\mcitedefaultmidpunct}
{\mcitedefaultendpunct}{\mcitedefaultseppunct}\relax
\EndOfBibitem
\bibitem[Johnson \emph{et~al.}(2014)Johnson, Wolk, Fournier, Sullivan, Weddle,
  and Johnson]{Communication2014johnson}
C.~J. Johnson, A.~B. Wolk, J.~A. Fournier, E.~N. Sullivan, G.~H. Weddle and
  M.~A. Johnson, \emph{J. Chem. Phys.}, 2014, \textbf{140}, 221101\relax
\mciteBstWouldAddEndPuncttrue
\mciteSetBstMidEndSepPunct{\mcitedefaultmidpunct}
{\mcitedefaultendpunct}{\mcitedefaultseppunct}\relax
\EndOfBibitem
\bibitem[Schr\"oder \emph{et~al.}(2022)Schr\"oder, Gatti, Lauvergnat, Meyer,
  and Vendrell]{Coupling2022schroder}
M.~Schr\"oder, F.~Gatti, D.~Lauvergnat, H.-D. Meyer and O.~Vendrell,
  \emph{arXiv:2204.03744}, 2022\relax
\mciteBstWouldAddEndPuncttrue
\mciteSetBstMidEndSepPunct{\mcitedefaultmidpunct}
{\mcitedefaultendpunct}{\mcitedefaultseppunct}\relax
\EndOfBibitem
\bibitem[Yang \emph{et~al.}(2020)Yang, Duong, Kelleher, and
  Johnson]{Capturing2020yang}
N.~Yang, C.~H. Duong, P.~J. Kelleher and M.~A. Johnson, \emph{Nat. Chem.},
  2020, \textbf{12}, 159--164\relax
\mciteBstWouldAddEndPuncttrue
\mciteSetBstMidEndSepPunct{\mcitedefaultmidpunct}
{\mcitedefaultendpunct}{\mcitedefaultseppunct}\relax
\EndOfBibitem
\bibitem[Dahms \emph{et~al.}(2016)Dahms, Costard, Pines, Fingerhut, Nibbering,
  and Elsaesser]{Hydrated2016dahmsa}
F.~Dahms, R.~Costard, E.~Pines, B.~P. Fingerhut, E.~T.~J. Nibbering and
  T.~Elsaesser, \emph{Angew. Chem. Int. Ed.}, 2016, \textbf{55},
  10600--10605\relax
\mciteBstWouldAddEndPuncttrue
\mciteSetBstMidEndSepPunct{\mcitedefaultmidpunct}
{\mcitedefaultendpunct}{\mcitedefaultseppunct}\relax
\EndOfBibitem
\bibitem[Carpenter \emph{et~al.}(2020)Carpenter, Yu, Hack, Dereka, Bowman, and
  Tokmakoff]{Decoding2020carpenter}
W.~B. Carpenter, Q.~Yu, J.~H. Hack, B.~Dereka, J.~M. Bowman and A.~Tokmakoff,
  \emph{J. Chem. Phys.}, 2020, \textbf{153}, 124506\relax
\mciteBstWouldAddEndPuncttrue
\mciteSetBstMidEndSepPunct{\mcitedefaultmidpunct}
{\mcitedefaultendpunct}{\mcitedefaultseppunct}\relax
\EndOfBibitem
\end{mcitethebibliography}

\end{document}